%% file: New_Paper.tex
\documentclass[a4paper,aps,prb,twocolumn,floatfix,showpacs,superscriptaddress,footnoteinbib]{revtex4}
\usepackage{graphics}
\usepackage{epsfig}
\usepackage{bbm}
\usepackage{times}
\usepackage{xcolor}   
\usepackage{amsfonts}
\usepackage{amsmath}

\usepackage{amssymb}
\usepackage{amsthm}
\usepackage{hyperref} 
\usepackage{wasysym}
\usepackage{bm}

\begin{document}

\title{Robust Majorana bound state in pseudo-spin domain wall of 2-D topological insulator
}

\author{Subhadeep Chakraborty}
\affiliation{Department of Physical Sciences, IISER Kolkata, Mohanpur, West Bengal 741246, India.}
\author{Vivekananda Adak}
\affiliation{Department of Physical Sciences, IISER Kolkata, Mohanpur, West Bengal 741246, India.}
\affiliation{Department of Physics, Korea Advanced Institute of Science and Technology, Daejeon 34141, Korea}
\author{Sourin Das}
\affiliation{Department of Physical Sciences, IISER Kolkata, Mohanpur, West Bengal 741246, India.}

\begin{abstract}
We investigate helical edge states (HES) emerging at the composite domain wall of spin and pseudo-spin degrees of freedom in a 2-D bulk governed by the Bernevig-Hughes-Zhang Hamiltonian which underwent quantum spin Hall to anomalous Hall transition. We numerically study the stability of Majorana bound state (MBS) formed due to proximity induced superconductivity in these helical edge states. We establish exceptional robustness of MBS  against moderate chemical potential or magnetic disorder owing to the existence of the simultaneous orthogonality between the right and the left moving modes both in spin and pseudo-spin space. Hence our proposal could pave the way to realizing robust Majorana bound state on 2-D platforms.
\end{abstract}

\maketitle

\textit{\underline{Introduction}:} Majorana bound states (MBS) \cite{alicea2012new, beenakker2013search, oreg2010helical} are exotic, self-conjugate excitations that can emerge at the boundaries of one-dimensional topological superconductors or in vortices of two-dimensional topological superconductors. They follow non-abelian statistics \cite{nayak19962n, nayak2008non, read2000paired, ivanov2001non, stern2004geometric, alicea2011non} and hold immense potential for application in quantum computing and topological quantum error correction \cite{nayak2008non, kitaev2003fault, alicea2011non, sarma2005topologically, stern2006proposed, zilberberg2008controlled, jiang2011interface, hassler2011top, flensberg2011non, clarke2011majorana, bonderson2011topological, bonderson2010implementing}. Kitaev's pioneering proposal for observing isolated MBS in one-dimensional spin-less p-wave superconductors \cite{kitaev2001unpaired} gives an unprecedented impetus to this direction of research. Subsequently, various approaches have been proposed to experimentally realize MBS in s-wave superconductor (SC) proximitized semiconductor nanowires subjected to Rashba spin-orbit coupling and Zeeman field \cite{lutchyn2010majorana, alicea2011non, danon2015interaction, rokhinson2012fractional, deng2012anomalous, oreg2010helical, stanescu2011majorana}.
 
 Detecting MBS in a transport setup hinges on the observation of smoking-gun evidence, such as the $2e^2/h$ zero-bias conductance peak \cite{mi2013proposal} and a 4$\pi$ Josephson effect \cite{fu2009josephson, kwon2004fractional}. Though there have been some encouraging developments in the nano-wire approach \cite{pikulin2021protocol}, a 2-D platform is desirable as it is well suited from the viewpoint of scalability. In recent times there has been some effort in exploring the potential of quantum spin-Hall (QSH) systems hosting helical edge states (HES) in proximity to an SC for realizing MBS \cite{hart2014induced}. However, it is also a challenging path due to the backscattering in HES in the presence of disorder \cite{beenakker2013search, maciejko2009kondo, maciejko2012kondo, tanaka2011conductance, lezmy2012single, budich2012phonon, schmidt2012inelastic, lunde2012helical, eriksson2012electrical, vayrynen2013helical, del2013backscattering, eriksson2013spin, altshuler2013localization, geissler2014random, kainaris2014conductivity, vayrynen2014resistance, pikulin2014interplay, dolcetto2016edge, kimme2016backscattering, hsu2017nuclear, hsu2018effects}. One promising strategy to address this challenge involves introducing generalized exchange potential \cite{liu2008quantum, li2013stabilization} into the QSH edge state such that it effectively pushes the two edge states (right and left movers) physically away from each other. Consequently, this approach can be used to drive the QSH state into a quantum anomalous-Hall(QAH) state, resulting in a chiral edge immune to backscattering.

\begin{figure}
 \centering
  \includegraphics[width=\columnwidth]{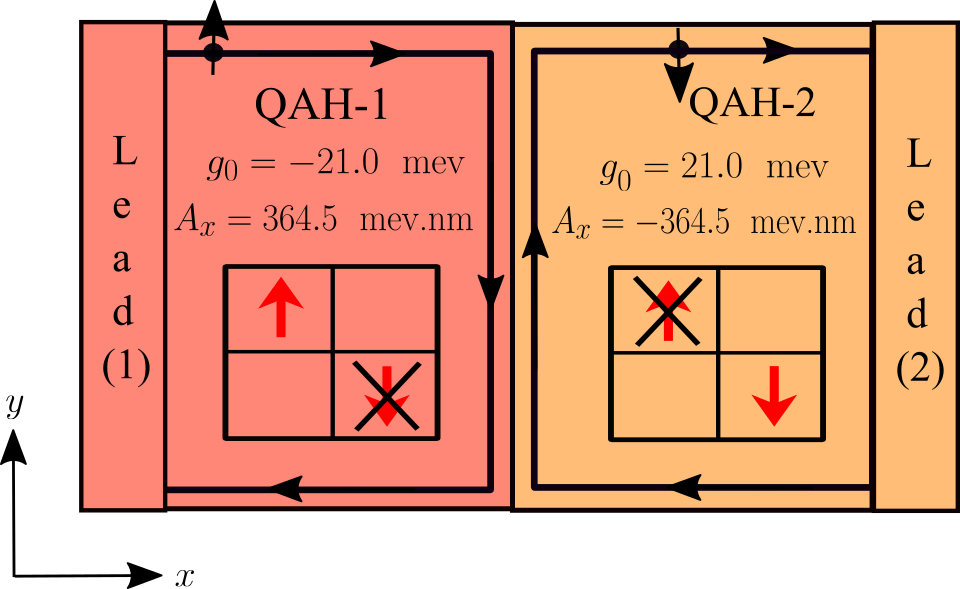}
  \caption{Schematic of the lattice model set-up employed for simulation. A QSH system of length $L$ and width $W$ is subjected to the influence of a spin and pseudo-spin domain wall (see Eq. \ref{bhzQAH1}), creating counter-propagating edge states at the domain wall with spin and pseudo-spin orthogonality. This gives rise to the emergence of a HES. The left and right regions of the domain wall are two oppositely spin-polarized QAH systems which are named ''QAH-1" and ``QAH-2" respectively.
 We attach two semi-infinite adiabatic leads, namely ``Lead-1" and ``Lead-2" to the left (``QAH-1") and right (``QAH-2") side of the system which follow the same model Hamiltonians as``QAH-1" and ``QAH-2" respectively.  $A_x$ represents the value of $A$ which appears in the $x$-hopping term of Eq. \ref{bhzQAH1}. }
 \label{schematic}
\end{figure}

In this article, we propose a two-dimensional set-up to realize robust MBS. We propose to achieve this by considering a QSH state split into two halves (see Fig. \ref{schematic}) such that each half is driven into a QAH state leaving behind a junction region hosting a helical edge with an interesting twist elaborated below. This could be achieved by driving the two halves of the 
QSH state to a spin-$\uparrow$ polarized QAH state and spin-$\downarrow$ polarized QAH system but with inverted pseudo-spin degrees of freedom \cite{roy2016pseudospin}, which results in the electrons in both regions flowing in a clockwise manner on the edge. Despite being in the immediate spatial neighbourhood of each other, these two chiral edges at the junction of the two QAH regions remain impervious to backscattering owing to pseudo-spin orthogonality in addition to the spin orthogonality.

We investigate the transport of electrons across the emergent HES  by employing quantum transport in 2-D lattice simulations using the Bernevig-Hughes-Zhang (BHZ) model \cite{bernevig2006quantum}.
The key results of our numerical simulations can be summarized as follows: (i) Backscattering of electrons in the emergent HES enjoys unprecedented protection against local disorder potential which breaks the spin or the pseudo-spin rotational symmetry individually (ii) The emergent HES when exposed to a proximity-induced superconductivity hosts MBS which show remarkable resilience against chemical potential or time-reversal symmetry breaking magnetic disorder of moderate strength (with respect to the bulk QAH gap).


\textit{\underline{The lattice model}}: To explore the transport through the emergent HES (discussed above), we perform a numerical tight-binding simulation of a two-dimensional topological insulator that accommodates chiral edge states at the boundary. This simulation is carried out using the KWANT package \cite{groth2014kwant}. We consider the square lattice model of a two-dimensional topological insulator given by the discretized version of the BHZ model \cite{chen2016pi, adak2024spin}.

 In Fig. \ref{schematic}, a schematic diagram of our transport set-up is presented. As elucidated before, we consider a QSH region of length $L$ and width $W$. This QSH region is described by the BHZ Hamiltonian, 

\begin{equation}
    H_{\rm BHZ} = -D k^2 + A k_x \sigma_z \tilde{\sigma}_x - A k_y \tilde{\sigma}_y + ({\cal M}-B k^2) \tilde{\sigma}_z,
 \label{BHZham}
\end{equation}

where $\sigma$ and $\tilde{\sigma}$ denote the Pauli matrices to describe the spins (up or down along $S_z$) and the orbitals degrees of freedom respectively and $A$, $B$, $D$ and ${\cal M}$ are material dependent parameters. We discretize the Hamiltonian in Eq.~\ref{BHZham} to obtain a tight-binding version on our square lattice where $k^2$, $k_x$ and $k_y$ can be approximated as $2 a^{-2}[2-\cos(k_x a)-\cos(k_y a)]$, $ a^{-1} \sin(k_x a)$, and $ a^{-1} \sin(k_y a)$ respectively, such that the tight-binding Hamiltonian reads \cite{chen2016pi, adak2024spin}
\begin{equation}
H_{\rm tb} = \sum_i (c_i^{\dagger} H_{i,i+a_x} c_{i+a_x} + c_i^{\dagger} H_{i,i+a_y} c_{i+a_y} + {\rm h.c.}) + c_i^{\dagger} H_{ii} c_i,
\end{equation}
where $c_i^{\dagger} \equiv (c^{\dagger}_{i,1,\uparrow},c^{\dagger}_{i,2,\uparrow},c^{\dagger}_{i,1,\downarrow},c^{\dagger}_{i,2,\downarrow})$ denotes the set of creation operators for the electrons in $1$ and $2$ orbital with $\uparrow$ and $\downarrow$ spins at site $i$ with coordinates $i=(i_x,i_y)$; $a_x = a(1,0)$ and $a_y = a(0,1)$ are the lattice vectors with $a$ being the lattice constant. As elucidated before, a QAH system can be obtained from the QSH system by inducing a topological phase transition driven by the application of an exchange field of the form $\sigma_z \tilde{\sigma}_z$ with strength $g_0$ in the BHZ Hamiltonian with $|g_0| > |\mathcal M|$ \cite{liu2008quantum, li2013stabilization}.  For $g_0 < \mathcal M$ ( assuming we are topological phase, i.e., ${\cal M}<0$ ), the spin polarization of the chiral edge at the boundary of the QAH region is spin-$\uparrow$ polarized whereas for $g_0 > - \mathcal M $, it is spin-$\downarrow$ polarized.



We impose a domain wall in the exchange field profile given by $g(x)=g_0\left[\Theta(L/2-x)- \Theta(x-L/2)\right] $ (with $|g_0|>|\mathcal{M}|$), which creates two chiral co-propagating edges with orthogonal spin-polarization at the domain wall. To make these edges counter-propagating like in a HES, we impose an additional domain wall in the parameter  $A\left[\Theta(L/2-x)- \Theta(x-L/2) \right]$ (see Eq. \ref{bhzQAH1}). Changing the parameter $A$ to $-A$ in the $x-$hopping inverts the propagation direction of electrons along the edge, which means in presence of this domain wall, electrons in the left and right half of the system flow with the same helicity (see App. \ref{App-A} for more details). This also gives rise to mutual orthogonality in pseudo-spin states of the two counter-propagating vertical edges at the domain wall which we will discuss in detail in the next section. So, the simultaneous presence of the domain walls in $g(x)$ and $A(x)$ creates orthogonality in spin and pseudo-spin states of the counterpropagating edge states at the domain wall and gives rise to an emergent HES with orthogonality in spin and pseudo-spin space. The tight-binding Hamiltonian in the whole region in the presence of the aforementioned domain walls can be written in a concise form,

\begin{align}
H_{ii} &= -\frac{4 D}{a^2} - \frac{4 B}{a^2} \tilde{\sigma}_z + {\cal M} \tilde{\sigma}_z \nonumber \\ &+ g_0 \left[ \Theta(i_{x_c}-i_x) -\Theta(i_{x}-i_{x_c}) \right] \sigma_z \tilde \sigma_z, \nonumber \\
H_{i,i+a_x} &= \frac{D + B \tilde{\sigma}_z}{a^2} + A\left[  \Theta(i_{x_c}-i_x) - \Theta(i_x-i_{x_c})
 \right]\frac{ \sigma_z \tilde{\sigma}_x}{2 i a}, \nonumber \\
H_{i,i+a_y} &= \frac{D + B \tilde{\sigma}_z}{a^2} + \frac{i A \tilde{\sigma}_y}{2 a}.
\label{bhzQAH1}
\end{align}

with $i_{x_c}$ being the lattice site corresponding to length $L/2$ along $x$-coordinate, so $i_{x_c} a =L/2$.   As can be seen in Fig. \ref{schematic}, the left and right halves of the system are named as ``QAH-1" and ``QAH-2" respectively. To facilitate reflectionless spin-polarized electron injection, we attach two semi-infinite QAH leads to the left and right of the system which are tagged as ``Lead-1" and ``Lead-2" leads respectively in Fig. \ref{schematic}.

\begin{figure}
 \centering
  \includegraphics[width=\columnwidth]{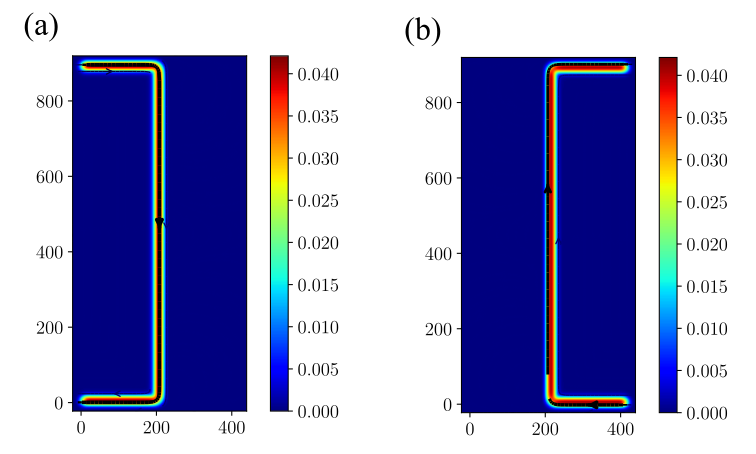}
  \caption{Plots depicting (a) probability current for spin-$\uparrow$ electrons injected from the ``Lead-1" and  (b) probability current for spin-$\downarrow$ electrons injected from the ``Lead-2" in absence of any disorder potential at the junction region of ``QAH-1" and ``QAH-2". The arrows in the plots denote the propagation direction of injected electrons which shows that the  spin-$\uparrow$ and spin-$\downarrow$ electrons injected from ``Lead-1" and ``Lead-2" are counter-propagating at the domain wall (junction region of ``QAH-1" and ``QAH-2"). The horizontal and vertical axes of the rectangular region represent the actual physical dimension of the system in nm.}
 \label{free_current}
\end{figure}

\begin{figure}
 \centering
  \includegraphics[width=\columnwidth]{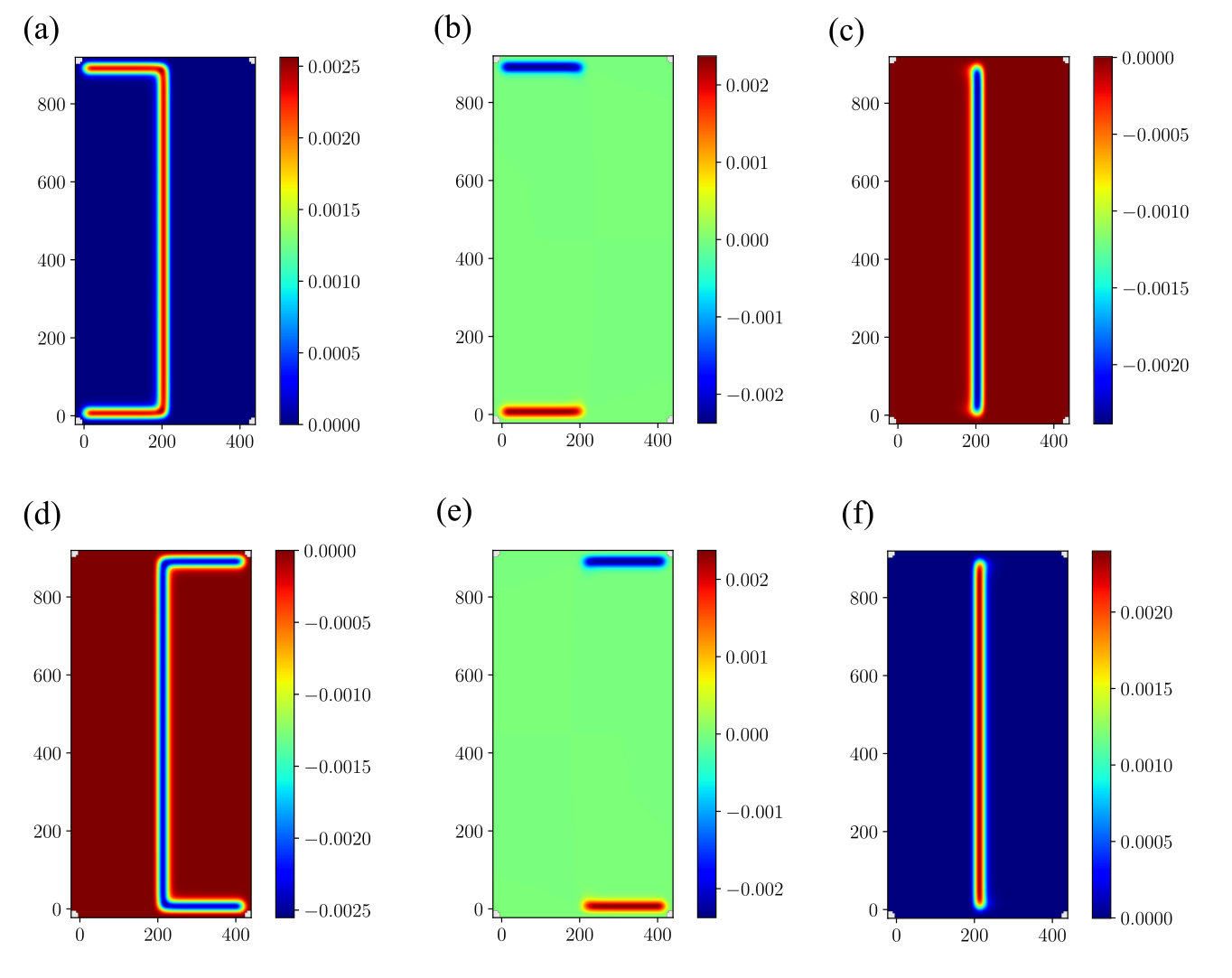}
  \caption{Plots depicting (a) spin ($z$-component)  (b) pseudospin ($x$-component) (c) pseudospin ($y$-component) densities of ``QAH-1" for injected spin-$\uparrow$ electrons from the ``Lead-1" and  (d) spin ($z$-component)  (e) pseudospin ($x$-component) (f) pseudospin ($y$-component) densities of ``QAH-2" for injected spin-$\downarrow$ electrons from the ``Lead-2" in absence of any disorder potential at the domain wall (junction region of ``QAH-1" and ``QAH-2"). The horizontal and vertical axes of the rectangular region represent the actual physical dimension of the system in nm.}
 \label{free_density}
\end{figure}

\textit{\underline{Transport across emergent HES}:} To simulate the lattice model set-up as described previously, we use the standard parameters for the HgTe/CdTe quantum wells that are~\cite{konig2007quantum} $A = \hbar v_F = 364.5$ nm meV, $B = -686$ ${\rm nm}^2$ meV, and ${\cal M} = -15$ meV while $D$ is set to zero to place the Dirac cone at zero energy, and the lattice constant is taken to be $a=3$ nm. The value of $g_0$ is taken to be $-21$ meV, such that ``QAH-1" is spin-$\uparrow$ polarized whereas ``QAH-2" is spin-$\downarrow$ polarized. Further, we consider $L=140a$ and $W=300a$.

Fig. \ref{free_current} shows the probability current of the spin-$ \uparrow $ and spin-$ \downarrow $  electrons injected from ``Lead-1" and ``Lead-2" respectively, where the arrows in the plots show the propagation direction of the injected electrons. This depicts that the spin-$ \uparrow $ and spin-$ \downarrow $ electrons injected from ``Lead-1" and ``Lead-2" are counter-propagating at the domain wall. We show the spin and pseudo-spin densities of electrons injected from the ``Lead-1" and ``Lead-2" in the absence of any local disorder potential in the emergent HES in Fig. \ref{free_density}. In Fig. \ref{free_density}-(a) and (d), we present the spin $S_z$ density of the propagating electrons in QAH-1 and QAH-2 when electrons are injected from the ``Lead-1" and ``Lead-2", respectively. These figures depict the independent existence of edge states in the ``QAH-1'' and ``QAH-2" regions, with ``QAH-1" exhibiting spin-$ \uparrow $ edge polarization and ``QAH-2" exhibiting spin-$ \downarrow $ edge polarization. 
This shows, that even when the wave functions for the two vertical edge states belonging to QAH-1 and QAH-2 are strongly overlapping in real space, they are not hybridizing owing to orthogonality discussed earlier.


In Fig. \ref{free_density}-(b), (c), (e), and (f), pseudo-spin polarization of the edge states in QAH-1 and QAH-2 are shown. The pseudo-spin polarizations of the top edges in QAH-1 and QAH-2 are oriented along the $-\hat x$ direction, while for the bottom edges, the pseudospin polarization is oriented along $\hat x$. The pseudo-spins of the vertical edges in the two QAH regions are $-\hat y$ and $\hat y$ polarized, respectively. Due to the inherent block-diagonal structure of a standard BHZ Hamiltonian in momentum space, the edge state vectors of the QAH-1 and QAH-2 edges can be elegantly expressed as the direct product of spin and pseudo-spin states, respectively\cite{roy2016pseudospin}. We denote the vertical edge states of QAH-$i$ at the domain wall as  $|V_i\rangle$ (for $i=1,2$) and building upon the preceding analysis, the spinors can be  expressed as (see Supplemental Material Section \ref{App-A}),

\begin{align}
    |V_1 \rangle \sim |\uparrow_z \rangle_s \otimes |\downarrow_y\rangle_p ,\nonumber \\
    |V_2 \rangle \sim |\downarrow_z \rangle_s \otimes |\uparrow_y\rangle_p
    , \label{QAH1spinor}
\end{align}




 where $s$ and $p$ denote spin and pseudo-spin spaces, respectively, $|\uparrow_i\rangle$ and $|\downarrow_i\rangle$ represent the eigenstates of the $i$-th Pauli matrix ($i=x, y, z$). These findings lead to a conclusion: (i) the spin polarizations of ``QAH-1" and ``QAH-2" are orthogonal, and (ii) the pseudo-spin polarizations of the vertical edges in ''QAH-1" and ''QAH-2" at the domain wall are also orthogonal.

To investigate the transport characteristics across the emergent HES, we carry out transport simulation such that  spin-$\uparrow$ electrons are injected from ``Lead-1" and we 
 observe complete electron-to-electron reflection; no electron transmission from ``Lead-1" to ``Lead-2" lead as expected due to the orthogonality present in both the spin and pseudo-spin states of two edge channels $|V_1\rangle$ and $|V_2 \rangle$ of the emergent HES. Also, it is evident that at the domain wall, applying any kind of edge local potential term of type $\sigma_x \tilde \sigma_x$, $\sigma_x \tilde \sigma_z$, $\sigma_y \tilde \sigma_x$, $\sigma_y \tilde \sigma_z$ that simultaneously couples both the spin and pseudo-spin states of $|V_1\rangle$ and $|V_2\rangle$ should results in the opening of a gap in the emergent HES by hybridizing these states. Consequently, spin-$\uparrow$ electrons injected from ``Lead-1" should transmit perfectly into ``Lead-2" as spin-$\downarrow$ electrons. On the other hand, any potential which couples to only 
spin or pseudo-spin state should be of no consequence. Transmission of electrons in the presence and absence of various edge potentials in the emergent HES obtained from our transport simulations is presented in a tabular form in Table. I in Supplemental Material Section \ref{App-B}, which supports the above discussed. In the above discussion, the applied potentials on the edge follow a spatial profile $\Theta(L/2-x) \Theta(x-(L/2-1))$, which means it is present only at the domain wall where $L$ is defined in terms of lattice spacing is given in the previous section. In what follows we study the transport across the emergent HES in the presence of a local superconducting pairing term.

\textit{\underline{Transport across superconductivity proximitized HES} :} On the right end of the ``QAH-1" region, for a region of dimension $5a \times 300a$, we apply a local superconducting pairing term $\Delta \tau_x$, (a term $\Delta \left[\Theta(L/2-x)\Theta(x-(L/2-5))\right] \tau_x$ is added to the onsite part) at the junction region, where $\tau_i$ ($i=x, y, z$) is the Pauli matrix operating in particle-hole space.  The lattice model Hamiltonian in presence of this paring term reads as,

\begin{align}
    H_{ii} &= -\frac{4 D}{a^2} \tau_z - \frac{4 B}{a^2} \tilde{\sigma}_z \tau_z + {\cal M} \tilde{\sigma}_z \tau_z \nonumber \\
    &+ g_0 \left[ \Theta(i_{x_c}-i_x) -\Theta(i_{x}-i_{x_c}) \right]\sigma_z \tilde{\sigma_z} \nonumber \\
    &+ \Delta \Theta (i_x-(i_{x_c}-5)) \Theta (i_{x_c}-i_x) \tau_x, \nonumber \\
    H_{i,i+a_x} &= \frac{D + B \tilde{\sigma}_z}{a^2} \tau_z \nonumber \\
    &+ A \left[  \Theta(i_{x_c}-i_x) - \Theta(i_x-i_{x_c})
 \right] \frac{ \sigma_z \tilde{\sigma}_x}{2 i a} \tau_z, \nonumber \\
    H_{i,i+a_y} &= \frac{D + B \tilde{\sigma}_z}{a^2} \tau_z + \frac{i A \tilde{\sigma}_y}{2 a} \tau_z.
	\label{bhzhamdelta}
\end{align}

\begin{figure}
 \centering
  \includegraphics[width=\columnwidth]{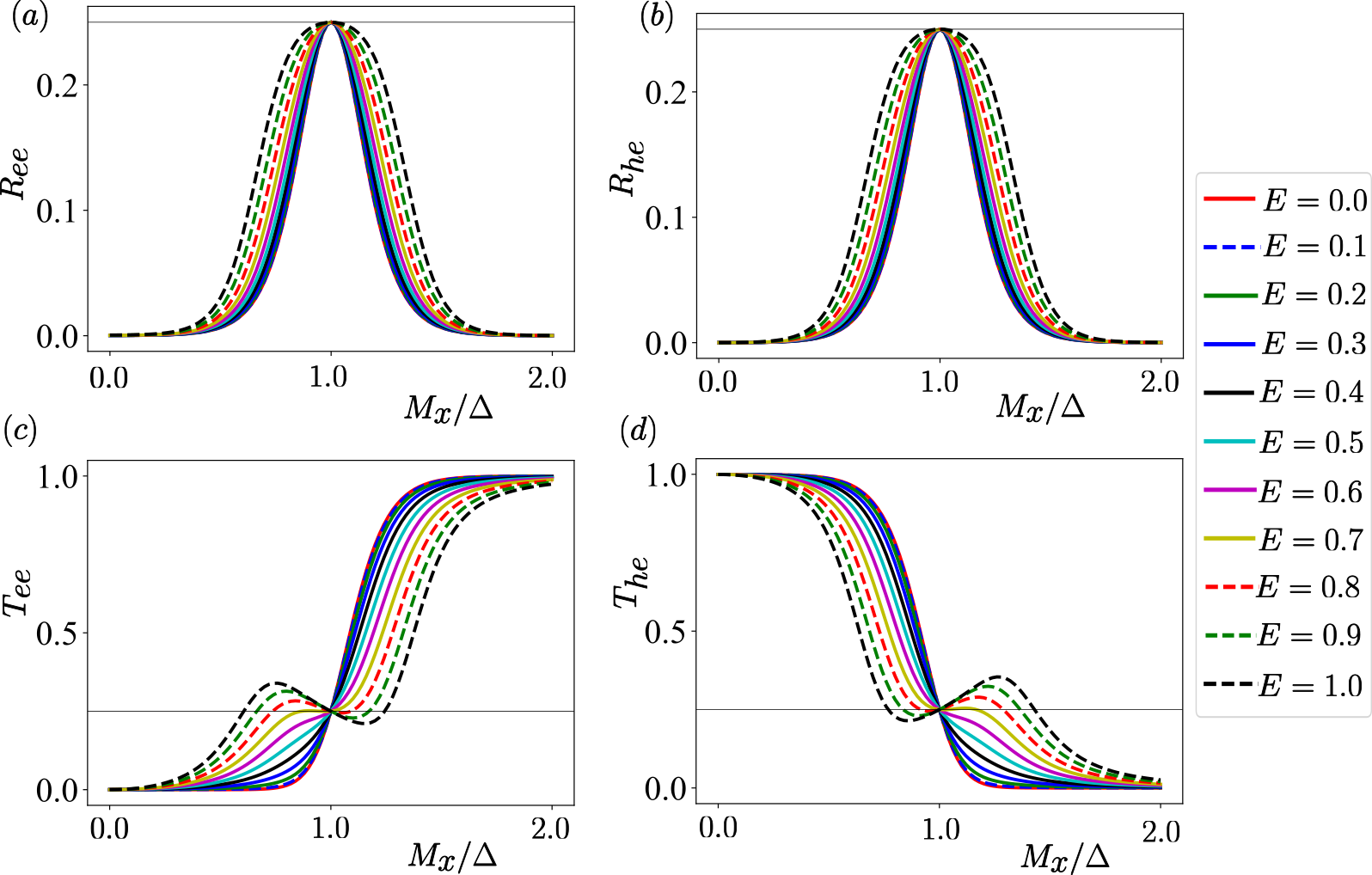}
  \caption{Plots depicting (a) $R_{ee}$ vs $M_x/\Delta$ (b) $R_{he}$ vs $M_x/\Delta$ (c) $T_{ee}$ vs $M_x/\Delta$ (d) $T_{he}$ vs $M_x/\Delta$ behaviour for electrons with incident energies $E$ (expressed in meV), keeping $\Delta$ fixed to $5$ meV. The potential ($M_x$) applied on the HES is of type $\sigma_x \tilde \sigma_x$. The faint black line represents the $0.25$ probability value. The non-local electrical conductance as measured in the ``Lead-2" is given by $G_{21}=(e^2/h) (T_{ee}-T_{he})$ \cite{fuchs2021crossed, zhang2019perfect, haugen2010crossed}. The plots show that at the critical point ($M_x=\Delta$, where the gap in the HES spectrum closes), the net electrical conductance measured in the ``Lead-2" is zero. As a consequence, zero average charge current is received at ``Lead-2", however, the average spin current in ``Lead-2" remains non-zero. }
 \label{Clean_Resonance}
\end{figure}

The Nambu basis consider here is given by, $\Psi = (\psi_{1\uparrow}, \psi^\dagger_{1\downarrow}, \psi_{2 \uparrow}, \psi^\dagger_{2\downarrow}, \psi_{1\downarrow}, -\psi^\dagger_{1\uparrow}, \psi_{2\downarrow} , -\psi^\dagger_{2 \uparrow})$ where $1,2$ represent the orbitals, and $\uparrow, \downarrow$ represent the spin indices along the $S_z$ direction. The plots depicting the transport of electrons across the superconductivity proximitized  HES as a function of the ratio of the superconducting gap ($\Delta$) and the edge potential ($M_x$) are shown in Fig. \ref{Clean_Resonance}, where $R_{ee}$, $R_{he}$. $T_{ee}$ and $T_{he}$ denote the electron-to-electron reflection, electron-to-hole reflection, electron-to-electron transmission, and electron-to-hole transmission probabilities respectively between the left and right lead. We define the local conductance in ``Lead-1"  as $G_{11}=(N_e + R_{he} - R_{ee})e^2/h$,  where $N_e$ is the number of channels for electron injection in ``Lead-1" and the non-local conductance in ``Lead-2" as $G_{21}=(T_{ee}-T_{he})$ \cite{fuchs2021crossed, zhang2019perfect, haugen2010crossed}. In the presence of the superconducting pairing term $\Delta$, the spin-$\uparrow$ electron injected from ``QAH-1" pairs up with another spin-$\downarrow$ electron from ``QAH-2" to form a cooper pair and the spin-$\downarrow$ electron leaves behind a spin-$\downarrow$ hole which is received at the ``Lead-2". Applying the edge local potential term $M_x \sigma_x \tilde \sigma_x$ in the region where local superconductivity has been induced, causes the emergent HES to coherently backscatter, thereby trying to open an insulating gap in the edge state. This gap competes with the superconducting gap, hence leading to suppression in $T_{he}$. When the edge local potential term becomes equal to the superconducting gap, reaching a gap-closing point (critical point), every type of possible transmission becomes equally probable, viz., $R_{ee}=R_{he}=T_{ee}=T_{he}=0.25$. A similar kind of transport where all transmission and reflection probabilities become equal to $1/4$ was also observed in a single channel $\mathcal{T}$-stub geometry strongly coupled to a superconductor at resonant energies of incidence\cite{das2009resonant}. At this critical point, the non-local electrical conductance  $G_{21}=(e^2/h) (T_{ee}-T_{he})$ \cite{fuchs2021crossed, zhang2019perfect, haugen2010crossed}, becomes zero which means zero average charge current is detected at ``Lead-2", while the average spin current in the lead should remain non-zero. As we cross the critical point, i.e., when the insulating gap exceeds the superconducting gap, we observe a rise in electron-to-electron transmission between the two leads. The Andreev conductance in ``Lead-1" (assuming electron injection from ``Lead-1") is expressed as $G=(1-T_{ee}+T_{he})e^2/h$ \cite{fuchs2021crossed, adak2022chiral, falci2001correlated} and the $2e^2/h$ Andreev conductance peak (when $T_{he}=1, T_{ee}=0$) at zero energy signifies the presence of MBS.

 These phenomena resemble the induction of superconductivity by proximity to a conventional helical edge, causing the edge spectrum to open a gap and host MBS at the end of proximitized region \cite{alicea2012new}. 
 However, the advantage of this emergent HES over conventional HES lies in the fact that the MBS hosted by the emergent HES remains remarkably robust against moderate chemical potential and time-reversal symmetry-breaking magnetic disorder, which we discuss in detail in the next section. We present the study of transport across the emergent HES in the presence of local superconductivity and edge local potentials at the emergent HES in a tabular form in Table. II in Supplemental Material Section \ref{App-B}.

\begin{figure}
 \centering
  \includegraphics[width=\columnwidth]{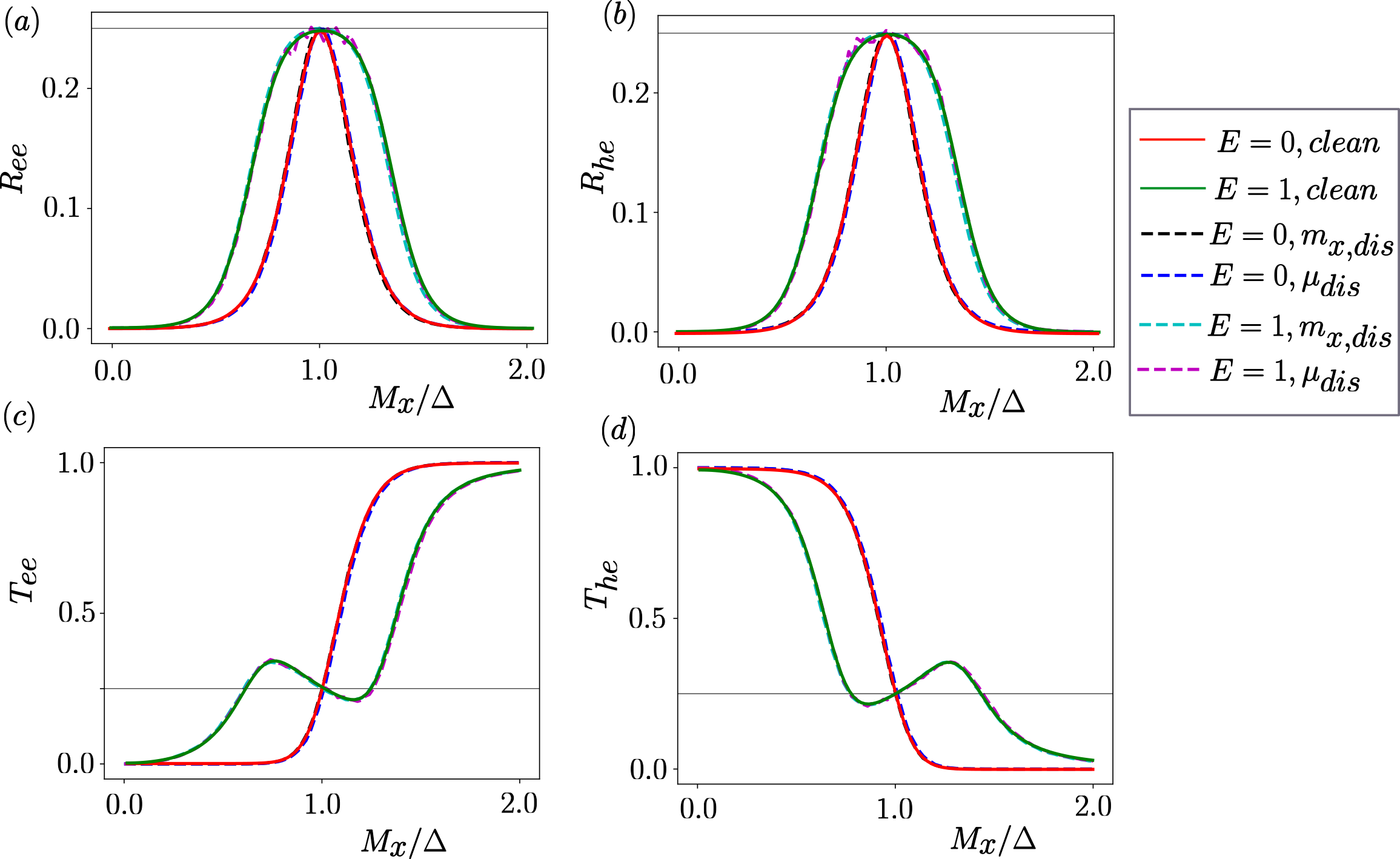}
  \caption{Plots depicting (a) $R_{ee}$ vs $M_x/\Delta$ (b) $R_{he}$ vs $M_x/\Delta$ (c) $T_{ee}$ vs $M_x/\Delta$ (d) $T_{he}$ vs $M_x/\Delta$ behaviour for electrons injected from ``Lead-1" with incident energies $E=0$ meV and $E=1$ meV in presence and absence of chemical potential ($\mu_{dis}$) and magnetic ($m_{x, dis}$) disorders with strengths taking random values from the range $[-2\Delta, 2\Delta]$ following a uniform distribution. In the plots, we keep fixed $\Delta=5$ meV and vary the edge local potential $M_x \sigma_x \tilde \sigma_x$ from 0 meV to 10 meV. The red, green, and blue dotted curves represent the transmission for clean, chemical potential, and magnetic disorder cases. The faint black line represents 0.25 probability value. The Andreev conductance in ``Lead-1" is given by $G=(e^2/h) (1-T_{ee}+T_{he})$.} 
 \label{Dis_reso}
\end{figure}

\textit{\underline{Transport in presence of disorder} :} We also numerically study the robustness of the MBS hosted by the emergent HES in the presence of chemical potential and magnetic disorder at the emergent HES. This is done by introducing additional onsite terms $\mu_i \tau_z$ and ${m_x}_{i} \sigma_x$ at the domain wall, with their strengths assuming random values distributed within the range of $[-2\Delta, 2 \Delta]$ meV ($\Delta$ fixed at 5 meV). By averaging over 10 samples, we present the results depicted in Figures \ref{Dis_reso}. These results depict that the transmission of electrons across the emergent HES remains remarkably robust, even in the presence of significant chemical potential and magnetic disorder. Notably, these disorders have minimal influence on the transmission, underscoring the protective role of simultaneous orthogonality in spin and pseudo-spin degrees of freedom in preserving the emergent HES, and hosting robust MBS in presence of a proximity induced superonductivity. This feature represents a distinct advantage of our proposal compared to conventional HES, where time-reversal symmetry-breaking disorder tends to induce backscattering.



\textit{\underline{Conclusion} :} In conclusion, our study offers compelling numerical evidence through a lattice model simulation of a 2-D  topological insulator, that the presence of the proposed domain wall  
leads to the emergence of an HES protected by spin and pseudo-spin orthogonality. This emergent HES in the presence of proximity induced superconductivity hosts MBS that exhibit remarkable resilience against chemical potential and magnetic disorders present in the system. As a result, this model presents a promising platform to realize disorder robust effective HES and Majorana zero modes.


\textit{\underline{Acknowledgement} :} We acknowledge Eytan Grosfeld for useful scientific discussions. SC acknowledges the Council of Scientific and Industrial
Research (CSIR), Govt. of India for financial support in
the form of a fellowship.


\pagebreak
\onecolumngrid
\input{ZAppendix}


\end{document}

%% file: ZAppendix.tex
\onecolumngrid

\appendix

\renewcommand\appendixname{Supplemental Material}

\section{Spin and pseudo-spin polarizations in presence of the domain walls} \label{App-A}

The pristine BHZ Hamiltonian in momentum space can be written down in a block-diagonal structure, and the edge states can be elegantly expressed as the direct product of spin and pseudo-spin states. So, the $2\times 2$ pseudo-spin block of the spin-$\uparrow$ polarized left half (``QAH-1") region  follows the Hamiltonian,

\begin{equation}
    h_u = (\mathcal{M} - Bk^2) \tilde{\sigma_z} + A_x k_x \tilde \sigma_x - A_y k_y \tilde \sigma_y
    \label{QAH_edge1}
\end{equation}

Solving for the vertical edge of the left half region (``QAH-1") at the domain wall, the edge state vector can be written down as \cite{qi2011topological}, 

\begin{equation}
    |V_1 \rangle = \begin{bmatrix}
        1 \\ 0
    \end{bmatrix}_s \otimes \begin{bmatrix}
        1 \\ -i
    \end{bmatrix}_p f(x) e^{-ik_y y}
\end{equation}

where $f(x)$ is a spatially decaying part which becomes zero at $x \rightarrow -\infty$. \\

In the right half (``QAH-2") region, we apply $A_x=-A, A_y=A$. So the pseudo-spin block of spin-$\downarrow$ polarized ``QAH-2" Hamiltonian becomes,

\begin{equation}
    h_d = (\mathcal{M} - Bk^2) \tilde{\sigma_z} + Ak_x \tilde \sigma_x - A k_y \tilde \sigma_y
    \label{QAH_edge2}
\end{equation}

The structure of $h_u$ and $h_d$ are identical, and that is why, the spin-$\uparrow$ electrons in ``QAH-1" and spin-$\downarrow$ electrons in ``QAH-2" flow with a similar helicity (in a clockwise fashion as shown in Fig. \ref{schematic} of the main text). So one can conveniently write down the vertical edge state vector of the right half region (``QAH-2") at the domain wall as,

\begin{equation}
    |V_2 \rangle = \begin{bmatrix}
        0 \\ 1
    \end{bmatrix}_s \otimes \begin{bmatrix}
        1 \\ i
    \end{bmatrix}_p f(-x) e^{ik_y y}
\end{equation}

where $f(-x)$ is a spatially decaying part which becomes zero at $x \rightarrow +\infty$. \\

As a result, at the domain wall, we have two oppositely spin-polarized counter-propagating edge states with orthogonal pseudo-spin polarization (see Fig. \ref{free_current} and Fig. \ref{free_density} of the main text) which gives rise to an emergent HES.

\section{Transport of electrons at zero energy across the emergent HES}\label{App-B}

Below we present our study of transport of electrons at zero energy across the emergent HES both in presence and absence of proximitized local superconductivity ($\Delta$) and edge local potentials ($M_x$) at the emergent HES.

\begin{center}\label{table1}
\begin{table}
	\begin{tabular}{||c  c  c c c ||} 
		\hline
		$Sl~no. $  & $M_x$   & $E$ & $R_{ee}$  & $T_{ee} $  \\ [1ex] 
		\hline
		1 & 0   & 0 & 1  & 0  \\ [1ex]
		\hline
		2 & 5 $\sigma_x$  & 0 & 1 &  0  \\ [1ex]
		\hline
		3 & 5  $\tilde \sigma_x$  & 0 & 1  & 0 \\ [1ex]
		\hline
		4 & $5 (\sigma_x + \tilde \sigma_x) $  & 0 & 1 & 0 \\ [1ex]
		\hline
            5 & 5 $\sigma_x \tilde \sigma_x $ & 0 & 0  & 1  \\ [1ex]
		\hline 
	\end{tabular}
    \caption{\label{demo-table} Our investigation of transport across the emergent HES, both in the presence and absence of various edge local potentials at the domain wall at zero energy ($E=0$ meV), is shown. We apply different types of edge local potentials $M_x$ at the domain wall and examine their effects on the transport of electrons along the emergent HES. Throughout our discussions, we represent particle-to-particle transmission from ``Lead-1" to ``Lead-1" as reflection and ``Lead-1" to ``Lead-2" as transmission. We express the electron-to-electron reflection probability and electron-to-electron transmission probability as $R_{ee}$ and $T_{ee}$ respectively. All relevant energy units are expressed in meV. The electrical conductance as measured in ``Lead-2" is given by $G=(e^2/h) T_{ee}$.}
\end{table}
\end{center}

\begin{center}\label{table2}
\begin{table}
	\begin{tabular}{||c c  c c c c c c||} 
		\hline
		$Sl~no. $  & $M_x$  & $\Delta$ & $E $ & $R_{ee}$ & $R_{he}$ & $T_{ee} $ & $T_{he}$ \\ [1ex] 
		\hline
		
		1 & 0  & 5 & 0 & 0 & 0 & 0 & 1 \\ [1ex]
		\hline
		2 & 5  $\sigma_x$ & 5 & 0 & 0 & 0 & 0 & 1\\ [1ex]
		\hline
		3 & 5  $\tilde \sigma_x$ & 5 & 0 & 0 & 0 & 0 & 1\\ [1ex]
		\hline
		4 & 5  ($\sigma_x+\tilde \sigma_x$) & 5 & 0 & 0 & 0 & 0 & 1\\ [1ex]
		\hline
		5 & 5  $\sigma_x\tilde \sigma_x$ & 5 & 0 & 0.25 & 0.25 & 0.25 & 0.25\\ [1ex]
		\hline
		6 & 2  $\sigma_x \tilde \sigma_x$ & 5 & 0 & 0.002 & 0.002 & 0 & 0.995\\ [1ex]
		\hline
		9 & 7  $\sigma_x \tilde \sigma_x$ & 5 & 0 & 0.017 & 0.017 & 0.966 & $  0$\\ [1ex]
		\hline
	\end{tabular}
 \caption{\label{demo-table} Our investigation of transport across the superconductor proximitized emergent HES, both in the presence and absence of various edge local potentials at zero energy ($E=0$ meV), is shown. We apply different kinds of edge local potential terms denoted as $M_x$ on the emergent HES and examine their effects on the transport of electrons. Throughout our discussions, we represent particle-to-particle transmission from ``Lead-1" to ``Lead-1" as reflection and ``Lead-1" to ``Lead-2" as transmission. We express the electron-to-electron reflection, electron-to-hole reflection, electron-to-electron transmission, and electron-to-hole transmission probabilities as $R_{ee}$, $R_{he}$, $T_{ee}$, and $T_{he}$, respectively. Assuming electron injection from ``Lead-1", the Andreev conductance in ``Lead-1" is expressed as $G=(e^2/h) (1 - T_{ee}+T_{he})$. All relevant energy units are represented in meV. At critical point $M_x=\Delta$, the average charge current detected in ``Lead-2" becomes zero, although the spin-current remains non-zero. }
\end{table}
\end{center}